\documentclass[showpacs,superscriptaddress,prl,twocolumn]{revtex4}%
\pdfoutput=1
\usepackage{amsmath}
\usepackage{graphicx}
\usepackage{bm}
\usepackage{amsfonts}
\usepackage{amssymb}%

\begin{document}
\title{Spin currents in rough graphene nanoribbons:\\
Universal fluctuations and spin injection}
\author{Michael~Wimmer}
\altaffiliation{These authors contributed equally to this work.}
\affiliation{Institut f{\"u}r Theoretische Physik, Universit{\"a}t
Regensburg, D-93040, Germany}
\author{\.{I}nan\c{c}~Adagideli}
\altaffiliation{These authors contributed equally to this work.}
\affiliation{Institut f{\"u}r Theoretische Physik, Universit{\"a}t
Regensburg, D-93040, Germany}
\author{Savas~Berber}
\affiliation{Institut f{\"u}r Theoretische Physik, Universit{\"a}t
Regensburg, D-93040, Germany}
\affiliation{Department of Physics and Astronomy, Michigan State University, East Lansing, Michigan 48824-2320, USA}
\author{David~Tom\'{a}nek}
\affiliation{Department of Physics and Astronomy, Michigan State University, East Lansing, Michigan 48824-2320, USA}
\affiliation{Institut f{\"u}r Theoretische Physik, Universit{\"a}t
Regensburg, D-93040, Germany}
\author{Klaus~Richter}
\affiliation{Institut f{\"u}r Theoretische Physik, Universit{\"a}t
Regensburg, D-93040, Germany}
\date{\today}


\begin{abstract}
We investigate spin conductance in zigzag graphene nanoribbons and
propose a spin injection mechanism based only on graphitic nanostructures.
We find that nanoribbons with atomically straight, symmetric edges
show zero spin conductance, but nonzero spin Hall conductance.
Only nanoribbons with asymmetrically shaped edges
give rise to a finite spin conductance
and can be used for spin injection into graphene. Furthermore, nanoribbons with rough
edges exhibit mesoscopic spin conductance fluctuations with a universal value
of $\mathrm{rms}\,G_\mathrm{s}\approx 0.4 e/4\pi$.
\end{abstract}

\maketitle

After their experimental discovery in 2004~\cite{Geim},
monolayers of graphite have attracted much experimental and
theoretical attention owing to their unusual band structure~\cite{GrapheneReviews}.
Graphene has also been suggested as a
good candidate for spin based quantum computing and
spintronics~\cite{Trauzettel}, as it is expected to have long
spin decoherence/re\-laxation times~\cite{MacDonald_dani06}. This
prospect led to the recent interest in generating and manipulating
net spin distributions in gra\-phene. Recently,
spin injection from ferromagnetic metal contacts into
graphene has been achieved~\cite{vanWees,Fuhrer,Hill,Shiraishi}.

Transport properties of graphene nanoribbons (GNR) are expected
to depend strongly on whether they have an armchair or zigzag edge~\cite{orient_cond}.
In GNRs with zigzag edges, transport is dominated by edge states
which have been observed in scanning tunneling microscopy~\cite{ES:exp}.
Moreover, owing to their high degeneracy, these states are expected
to be spin polarized~\cite{FujitaJPSJ}, making
zigzag GNRs attractive for spintronics~\cite{Son06}. In addition,
edge states are expected to occur also in nanoribbons with other
edge orientations~\cite{Akhmerov}.
Recently, the first transport experiments have been performed in
narrow ribbons of graphene~\cite{Kim}, albeit with not well defined edges.
Recent theoretical work focused on charge transport through rough GNRs~\cite{Blanter_CastroNeto},
but spin transport properties have not been explored yet.

In the present work, we focus on spin transport in GNRs with rough zigzag edges.
Ideal zigzag GNRs are not efficient spin injectors
due to the symmetry between the edges with opposite magnetization.
In order to obtain net spin injection, this symmetry must be broken.
Existing proposals to achieve this require very large transverse electric fields~\cite{Son06}.
We sidestep this difficulty by showing that edge imperfections (such as vacancies),
which usually cannot be avoided experimentally, break the symmetry between
the edges and lead to a finite spin conductance of the GNR. Thus,
rough zigzag GNRs can be used as spin injectors or detectors in graphene spintronics.
%to inject spins, or can act as a spin detector.

\begin{figure}[ptb]
\includegraphics[width=0.95\columnwidth]{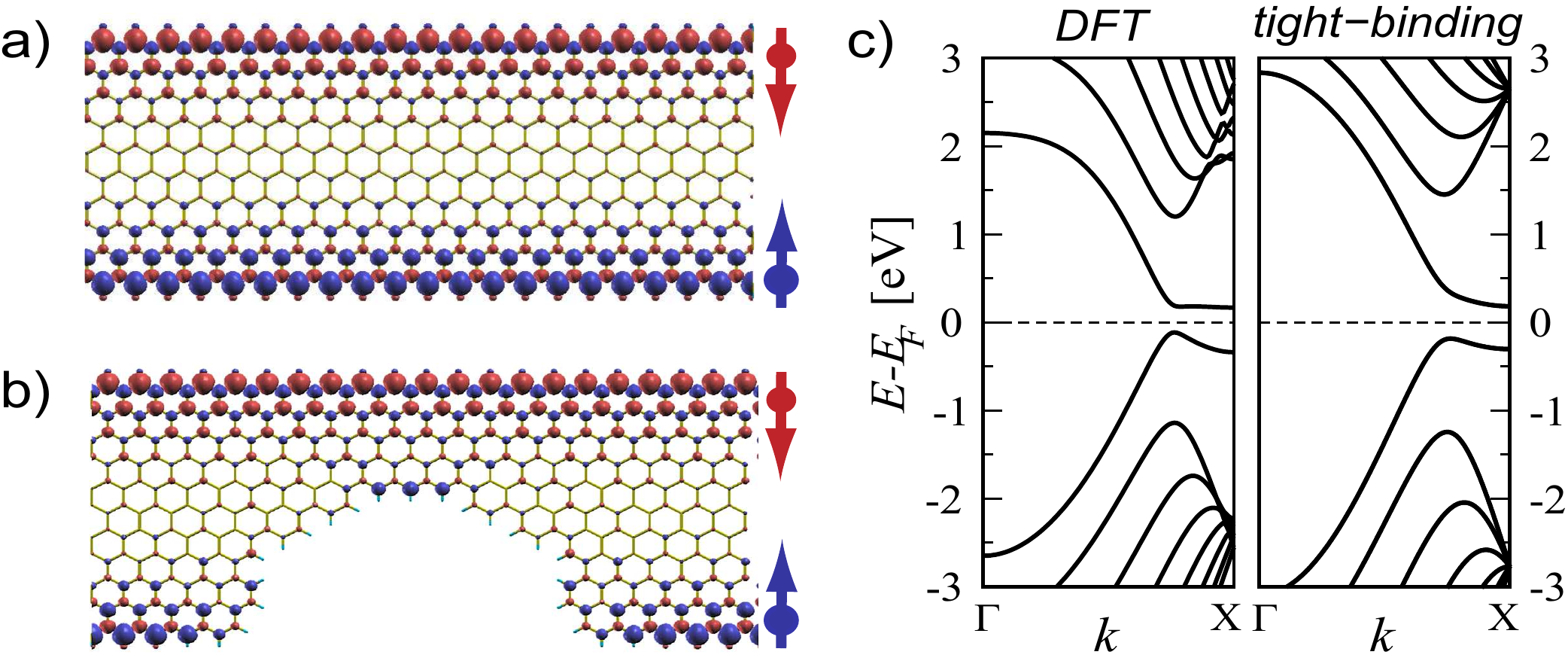}
\caption{(color online) Ground state spin
density for (a) an ideal and (b) an imperfect zigzag GNR. Blue
(red) corresponds to up (down) spin density. (c)~Band structures of
an ideal GNR obtained from DFT and tight-binding approaches.}
\label{FIG:GNR_band}%
\end{figure}

We start with a description of the electronic ground state properties
of the zigzag GNR, which captures the
essential physics relevant to spin transport, given by
the single band tight-binding Hamiltonian~\cite{FujitaJPSJ}
\begin{equation}\label{EQ:MF_hamiltonian}
H=\sum_{ij,s} t_{ij} c^\dag_{i,s} c^{\phantom \dag}_{j,s}
+ \sum_{i,s,s'}\mathbf{m}_i\cdot c^\dag_{i,s} \boldsymbol{\sigma}_{s,s'}
c^{\phantom \dag}_{i,s'}\,.
\end{equation}
Here $t_{ij}=t$ if $i$ and $j$ are nearest neighbors, $t_{ij}=t'$ if $i$ and $j$ are next nearest
neighbors~\cite{TBGraphene},
and $\boldsymbol{\sigma}$ are the Pauli matrices corresponding to
the spin degree of freedom. The local magnetization $\mathbf{m}_i$
can be obtained from the self consistency condition or
\emph{ab initio} calculations.

Our \emph{ab initio} results, obtained using the spin-polarized
density functional formalism (DFT)~\cite{Siesta},
agree with the reported finding \cite{FujitaJPSJ, Okada, Son06}
that the local magnetization is staggered in the electronic
ground state, as shown in Fig.~\ref{FIG:GNR_band}(a).
At zero doping the antiferromagnetic (AF) ordering generates a
gap in the single particle spectrum. We now dope the GNR in order to move into a regime with open conduction channels.
This can be achieved in practice by applying a gate voltage or chemical doping.
Our DFT results indicate that a finite
amount of doping reduces the AF gap and the local magnetization, but does not destroy the AF ordering.
We obtain the critical value of this doping as ${\approx}0.5$ electrons
(${\approx}0.4$ holes) per zigzag edge atom. Furthermore, our DFT calculations show that not only
perfect, but also rough zigzag ribbons exhibit spin
polarization (Fig.~\ref{FIG:GNR_band}(b)). In addition, the formation of
multiple spin domains at zigzag edges is energetically prohibitive.
In summary, our DFT calculations show that it is possible
(i)~to dope the GNR to make them conductive and
(ii)~to introduce disorder at the edges while retaining the magnetic ordering.

\begin{figure}[ptb]
\mbox{ \includegraphics[width=0.43\columnwidth]{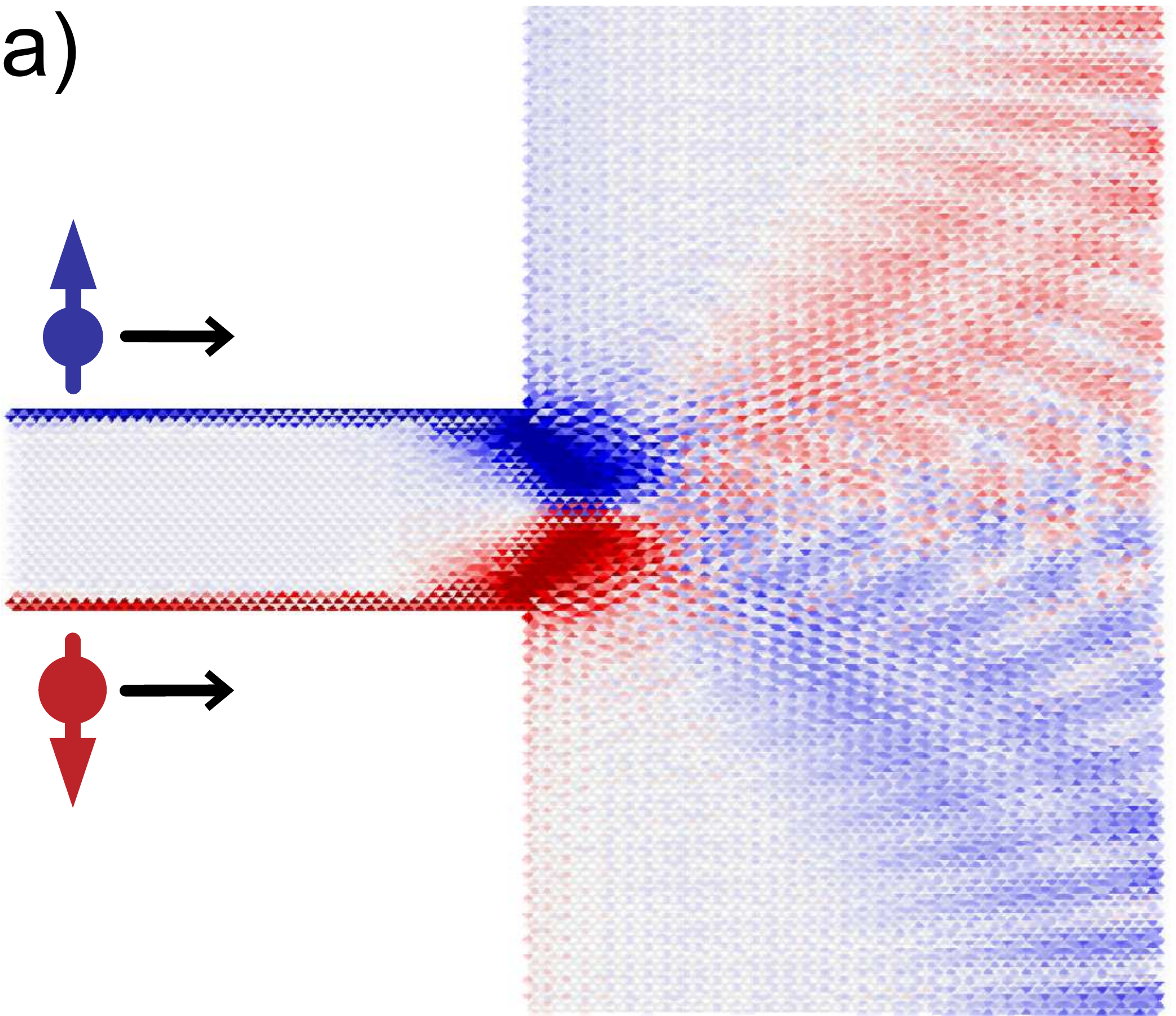}}
\hskip0.02\columnwidth
\mbox{\includegraphics[width=0.43\columnwidth]{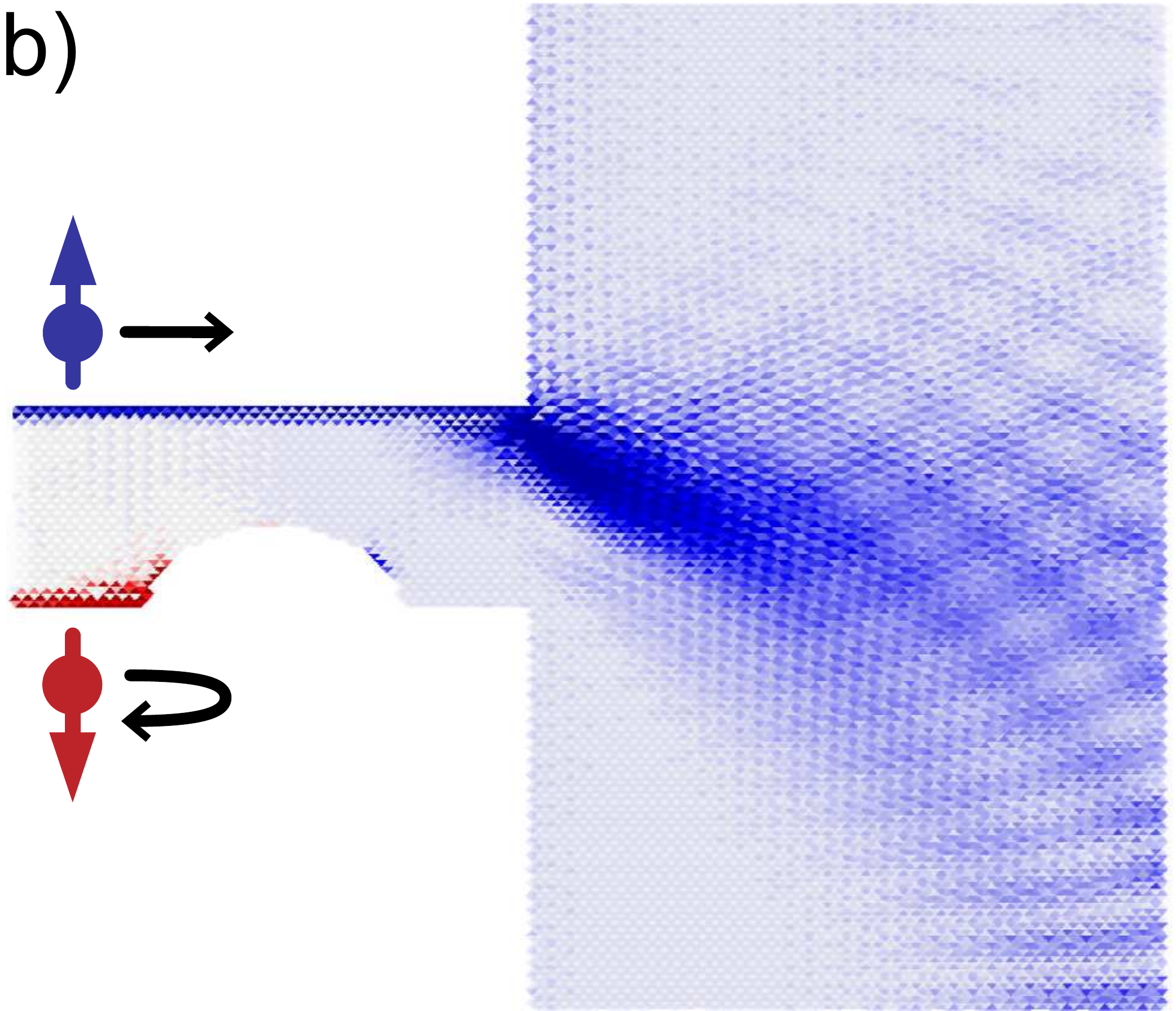}}
\caption{ (color online) Spin injection profile from (a) an ideal
GNR and (b) a GNR with a distorted edge into a region of n-doped
graphene. Nonequilibrium densities for spin up (down) electrons
are shown in blue (red). }
\label{FIG:Ideal_GNR}%
\end{figure}

Next, we further simplify the mean field description of Eq.~(\ref{EQ:MF_hamiltonian})
by ignoring the variation of $\mathbf{m}_i$ within
a sublattice. A spatial dependence of $\mathbf{m}_i$ changes the amount of band
% -bending,
dispersion, modifying the energy window, within which the
transport predominantly involves the edge states. This leads to
the single particle Hamiltonian
\begin{equation}
H_{mf}=\epsilon(k)\tau_1 +\Delta(k)\tau_2 + A(k) \mathrm{I}+\mathbf{m}\cdot \boldsymbol{\sigma} \tau_3,
\label{EQ:SP_hamiltonian}
\end{equation}
where $\epsilon(k)$, $\Delta(k)$ and $A(k)$ are obtained by Fourier transforming Eq.~(1), and
$\tau_i$ are the Pauli matrices corresponding to
pseudospin(sublattice) degrees of freedom~\cite{Haldane}.
%The homogeneous antiferromagnetic
The AF exchange field $\mathbf{m}$ is obtained by
fitting the band structure to DFT results (see Fig.~\ref{FIG:GNR_band}(c)).

In the following, we focus on
% the
transport properties of the GNR. We work in the linear response
regime so that all the transport properties
of the GNR are specified by the effective single-particle Hamiltonian~(\ref{EQ:SP_hamiltonian}).
The spin conductance~\cite{SCunit} of a GNR is given by
%\begin{equation}
$ G_s
%=(\hbar/2 e)(G_{\uparrow}-G_{\downarrow})
=(e/4\pi)(T_{\uparrow}-T_{\downarrow}),
$
%\end{equation}
where
%$G_{\uparrow(\downarrow)}$ is the conductance and
$T_{\uparrow(\downarrow)}$ is the transmission probability for
spin up (down). The conducting channels with energies closest to
the Fermi energy of the undoped system reside on a single
sublattice and are fully spin polarized owing to the staggered
magnetization.  These states are extended along the ribbon axis,
but localized near the (zigzag) edges, with the spin up channel
localized at one edge and the down channel on the opposite edge.
The transverse localization length of these states depends on
their Fermi momentum $k_F$ that may be modified by shifting the
Fermi energy $E_F$. As one moves away from the $\mathrm{X}$ point,
the transverse localization length increases as
$\lambda_\mathrm{edge}\approx-a/\ln(2\cos(k_F a/2))$, where
$a=2.46~\text{\AA}$ is the hexagonal lattice
constant~\cite{FujitaJPSJ}. Owing to the spatial separation of the
edge states, the scattering of spin up and spin down carriers
occurs only at the edge, where they reside, and is unaffected by
the opposite edge. Distinguishing a left (l) and right (r) edge of
the nanoribbon, we approximate $T_{\uparrow(\downarrow)}$ by
$T_{\rm l(r)}$, where $T_{\rm l(r)}$ is the transmission
probability of the corresponding edge state, assuming the opposite
edge is not disordered. The transport properties of the zigzag GNR
are thus essentially those of two independent wires, oppositely
spin polarized and connected in parallel between the reservoirs.
We note that previous studies of edge state
transport~\cite{MunozNikolic} assumed vanishing next nearest
neighbor hopping $t'$, and obtained results in apparent
contradiction to the picture presented above: If $t'$ were zero,
the charge density would be localized at the edges, but the
current density would be extended through the GNR. This leads to
the incorrect conclusion that edge states would scatter equally
from impurities at \emph{both} edges. In reality, the edge states
show non-zero dispersion (such as due to $t'\neq 0$). In this
case, the current flow is also localized at the
edges~\cite{ourselves} validating the two-wire model, as we show
below.

For an ideal, impurity-free GNR, we have $T_{\rm l}=T_{\rm r}$,
which leads to vanishing spin conductance.
%-------new------------
This is confirmed
by quantum transport simulations~\cite{recursiveAlgo1} and an illustrative
example is shown in Fig.~\ref{FIG:Ideal_GNR}(a): Both edge channels
transmit equally.
However, as the edge states enter the bulk graphene, they are deflected:
In the GNR, the pseudospin is predominantly in $z$-direction
and tied to the electron spin, whereas in the bulk
pseudospin is in-plane and tied to the current direction.
At the interface, the $z$-component splits into states
with positive and negative velocity perpendicular to the
boundary. The state with velocity towards the boundary is
scattered~\cite{ourselves} and thus,
upon entry, states at opposite edges (which carry opposite spins) deflect
in opposite directions, leading to a finite spin Hall conductance
(Fig.~\ref{FIG:Ideal_GNR}(a)).
Finite spin conductance can be obtained, however, for imperfect
GNRs: An obstacle scatters the spin channel localized at the same
edge more effectively, leading to a non-vanishing spin conductance
and spin injection (Fig.~\ref{FIG:Ideal_GNR}(b)). Whereas the
efficiency of the spin Hall effect is limited by the mean free path,
and thus ballistic microstructures are needed to observe it, the
efficiency of spin injection with edge defects is limited only by
the spin relaxation length
and can be used to inject spins into diffusive systems.

\begin{figure}[t]
\includegraphics[width=0.90\columnwidth]{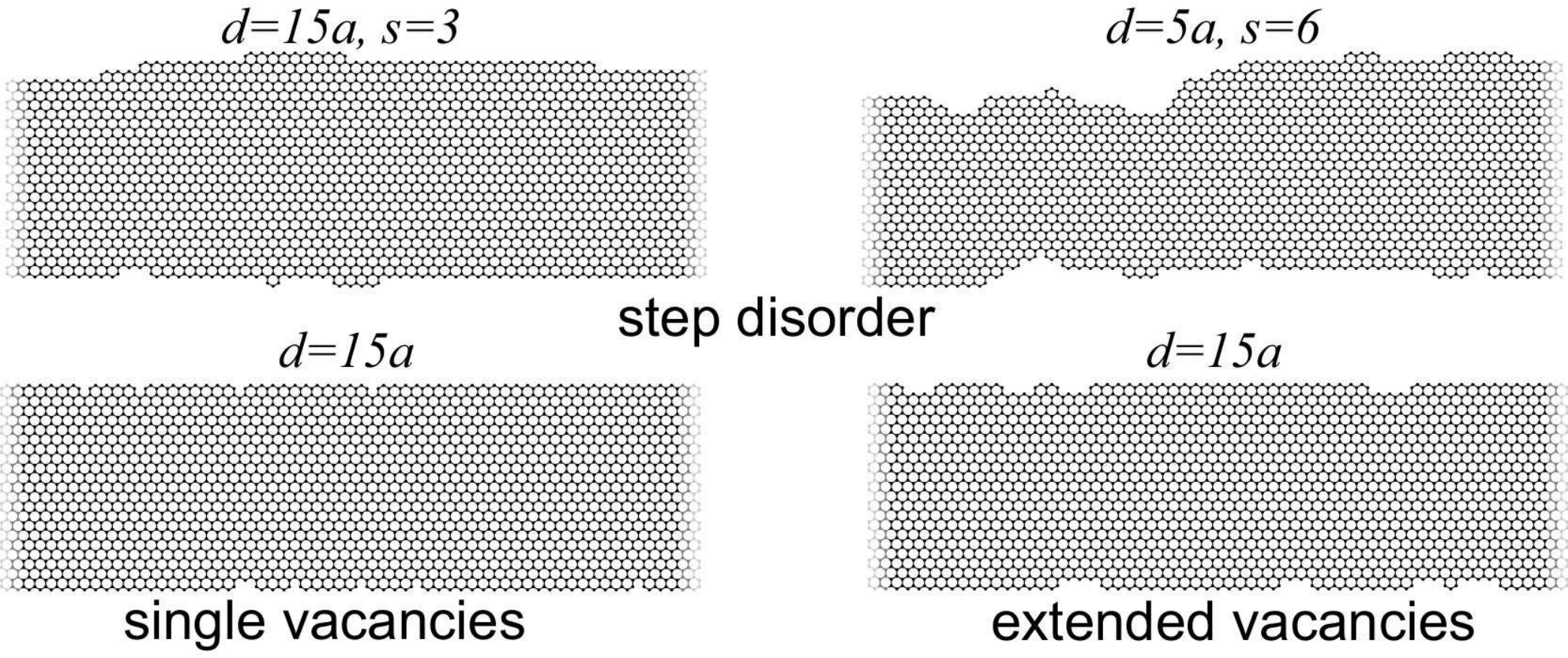}
\caption{ Step disorder: edge disorder created by a random walk,
where the width of the nanoribbon is changed by one hexagon at
every step. Steps are made with probability $a/d$ and the maximum
deviation of the width is $\leq s$ hexagons. Single vacancies:
edge atoms are removed randomly with the probability $a/d$.
Extended vacancies: similar to single vacancies, but also
neighboring edge atoms are removed.
%In all cases the disorder on the two edges is uncorrelated.
}
\label{FIG:disordermodels}
\end{figure}

From an experimental perspective, unless the GNRs are specifically
fabricated with edges of different roughness, the average
conductance of both spin channels is equal, quenching the
ensemble-averaged spin conductance. Yet, in the mesoscopic regime,
sample-to-sample fluctuations of $T_\mathrm{\uparrow,\downarrow}$
%in the conductance of the left and right channels
lead to a non-vanishing variance of the spin conductance. In the two-wire model
we have
%\begin{eqnarray}
\begin{equation}
\mathrm{Var}\,G_\mathrm{s}=\left(\frac{\hbar}{2e}\right)^2\mathrm{Var}\,G_\mathrm{tot}
=\left(\frac{e}{4\pi}\right)^2(\mathrm{Var}\,T_\mathrm{l}+
\mathrm{Var}\,T_\mathrm{r})\,. \nonumber
\end{equation}
%\end{eqnarray}
Treating both edges as one-dimensional wires, we map the transport problem onto that of a disordered
1D chain. Transmission eigenvalue statistics in 1D disordered chains is known to be described by
the Dorokhov-Mello-Pereyra-Kumar (DMPK) equation~\cite{DMPK}.
Using the full distribution function of
resistance~\cite{GertsenshteinBeenakkerMelsen},
we find that the universal maximum
value of the root mean square (rms) spin conductance $\mathrm{rms}\,G_s=\sqrt{\mathrm{Var}\,G_s}
\approx 0.4 e/4\pi$.
In order to demonstrate this universality,
we investigate GNRs of different length $L$ and width $W$ and various models of edge disorder (see
Fig.~\ref{FIG:disordermodels}).

\begin{figure}[ptb]
\includegraphics[width=0.92\columnwidth]{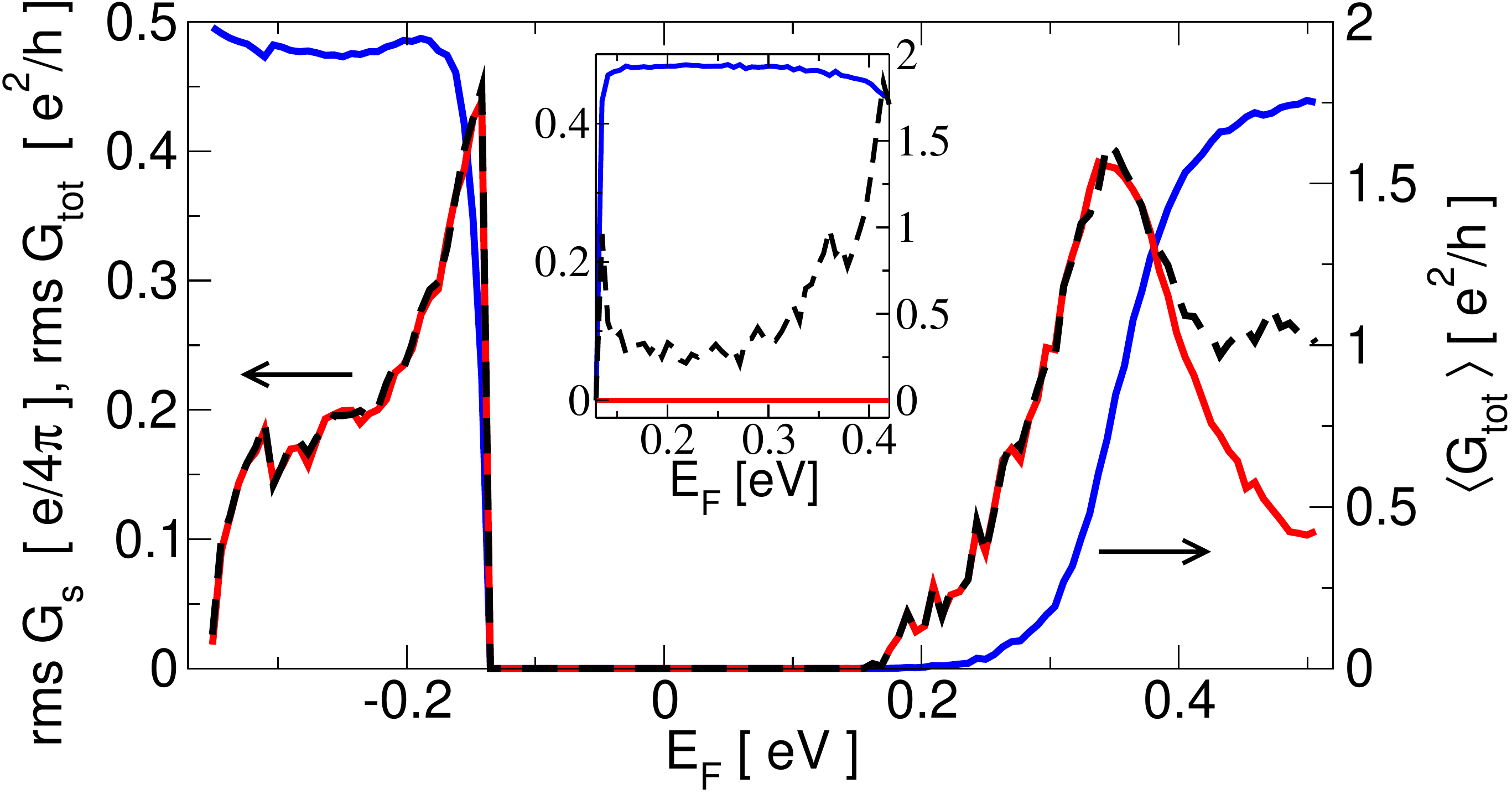}
\caption{ (color online) Average total conductance, $\langle
G_\mathrm{tot}\rangle$ (blue solid line), rms of the total
conductance, $\mathrm{rms}\,G_{tot}$ (black dashed line), and rms
of the spin conductance, $\mathrm{rms}\,G_\mathrm{s}$ (red solid
line), as a function of $E_F$ ($E_F=0$ is chosen to correspond to
zero gate voltage). The data were averaged over 1000
configurations of single vacancies with $d=40a$ and $L=800a$. For
comparison, the inset shows the same quantities for for the
singular case of $t'=0$. In this situation, the spin conductance
and its fluctuations vanish completely. } \label{FIG:example}
\end{figure}

First, we focus on dilute disorder, where the average distance between scatterers $d \gg a$.
The typical behavior of charge and spin conductances (average, fluctuations) is shown in
Fig.~\ref{FIG:example}. We first note that over the whole energy region, where the edge states are
present, $\frac{\hbar}{2e}\,\mathrm{rms}\,G_\mathrm{tot} \approx \mathrm{rms}\,G_\mathrm{s}$,
confirming the validity of the two-wire model. As the Fermi level is raised by gating or doping,
the relevant states are
extended and feel both edges. Then, the assumption of uncorrelated
channels breaks down, and $\frac{\hbar}{2e}\,\mathrm{rms}\,G_\mathrm{tot} > \mathrm{rms}\,G_\mathrm{s}$.

For an n-type GNR, when the Fermi level is near the band edge,
the states at $E_F$ are localized and both the average conductance
and the fluctuations are suppressed exponentially. Raising $E_\mathrm{F}$,
we observe in Fig.~\ref{FIG:example} a crossover to the ballistic regime, where the
conductance rises up to the quantum limit of conductance $2 e^2/h$.
Correspondingly, we see a maximum in the conductance fluctuations before they vanish
again in the ballistic regime.

The average/fluctuations of the conductances of a p-doped GNR are different from an
n-doped one, but a description based on the DMPK equation holds well for either case.
The scattering strength
of impurities depends on the overlap of the impurity
potential with the unperturbed channel wavefunction and therefore on
$\lambda_\mathrm{edge}=\lambda_\mathrm{edge}(E_\mathrm{F})$.
In the n-doped GNR, there is one channel whose momentum is a
monotonic function of $E_F$.
On the other hand, in the p-doped GNR, due to the band
% bending
dispersion (Fig.~\ref{FIG:GNR_band}), there are two channels: One
localized near the edge, the other extended further into the
ribbon, but still with a considerable density at the edge.
Lowering $E_F$ thus localizes one state even more towards the
edge, whereas the other state spreads out, making the density more
uniform. This leads to different functional dependences of the
localization length on the Fermi energy for n- and p-doped
ribbons.

\begin{figure}[ptb]
\includegraphics[width=0.83\columnwidth]{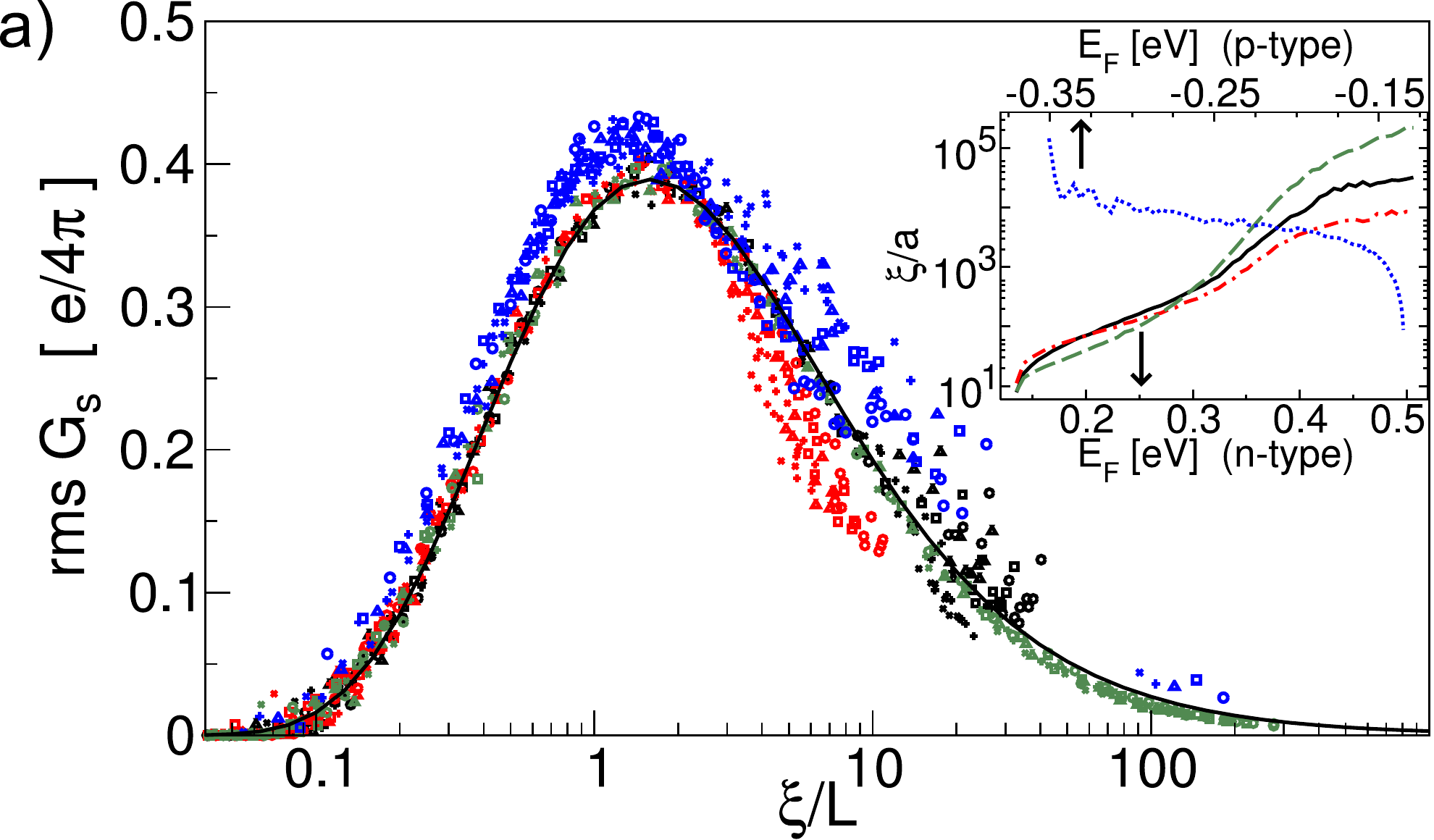}
\includegraphics[width=0.83\columnwidth]{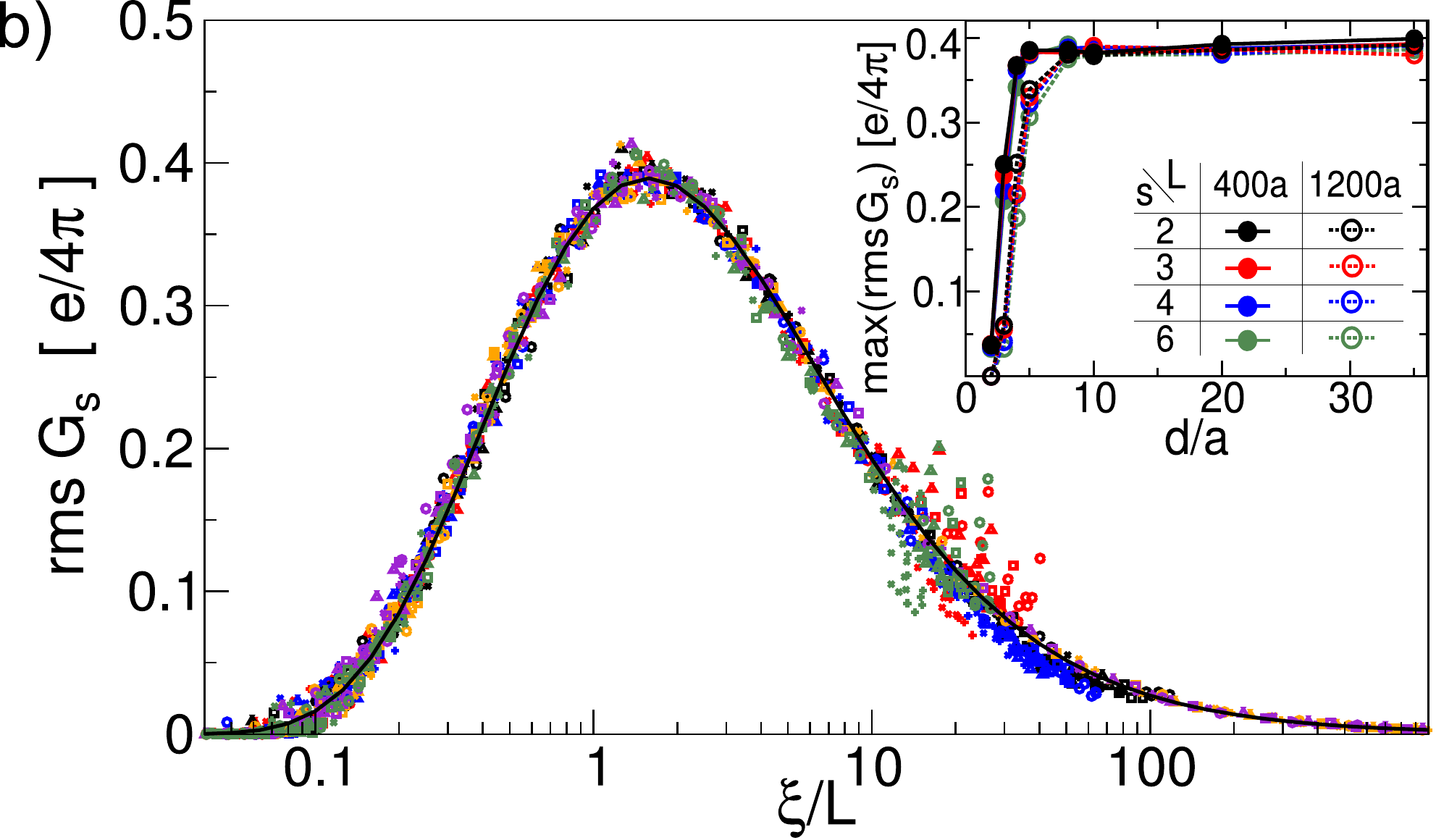}
\caption{(color online) Spin conductance fluctuations:
(a) $\mathrm{rms}\,G_\mathrm{s}$ as a function of $\xi/L$
for n- and p-doped graphene:
step disorder for n-type, $d=20a$, $s=3$
(black), single vacancies for n- and p-type, $d=40a$ (red and blue,
respectively) and
extended vacancies for n-type, $d=30a$ (green).
Inset: $\xi/a$ as a function of $E_F$ for different disorder models
(colors as in the main panel).
(b) $\mathrm{rms}\,G_\mathrm{s}$ as a function of $\xi/L$ for step disorder
in n-doped graphene:
$d=20a$ and $s=3$ (red; orange for $W=92a/\sqrt{3}$ ),
$d=35a$ and $s=2$ (black),
$d=35a$ and $s=6$ (blue; violet for $W=92a/\sqrt{3}$),
$d=20a$ and $s=6$ (green).
Inset: maximum value of $\mathrm{rms}\,G_\mathrm{s}$
as a function of $d/a$ for the step
disorder models.
In both (a) and (b), the solid line corresponds to the DMPK prediction.
The data is shown for GNR lengths $L=800a$ ($\bigcirc$),
$1000a$ ($\Box$), $1200a$ ($\triangle$), $1400a$ ($+$), and $1600a$ ($\times$),
width $W=32/\sqrt{3}a$ unless specified otherwise.
The $\mathrm{rms}\,G_\mathrm{s}$ is estimated from 1000 ($W=32a/\sqrt{3}$)
and 750 ($W=92a/\sqrt{3}$) disorder configurations.
}
\label{FIG:universality}
\end{figure}

In order to compare n- and p-doped ribbons as well as different disorder models,
we extract the energy dependence of the longitudinal (transport) localization length
$\xi(E_\mathrm{F})$ from $\exp(\langle ln(G_{\uparrow/\downarrow}(E_\mathrm{F},L) \rangle)=\exp(-2L/ \xi)$~\cite{Anderson, Beenakker},
as shown in the inset of Fig.~\ref{FIG:universality}(a).
In Fig.~\ref{FIG:universality}(a) we show $\mathrm{rms}\,G_\mathrm{s}$ as a function of $\xi/L$
for all three disorder models (see Fig.~\ref{FIG:disordermodels}) with different values of
$d$ and a wide range of ribbon lengths $L$. The data collapse onto a single curve, demonstrating the
universality of the spin conductance fluctuations (SCF), independent of the particular type
of edge disorder. Slight deviations from this universality can be observed in
Fig.~\ref{FIG:universality}(a), in the ballistic
regime for the special case of single vacancies. In this case,
the system reaches the ballistic limit only for high
Fermi energy values, where the two-wire model breaks down.
The rms spin conductance of the n-doped GNR agrees very well with the results obtained from the
DMPK equation.
For the p-doped ribbon, where there are two conducting channels, we see a small increase in the
rms conductance, presumably due to the crossover to a multi-channel quasi-1D wire,
where $\mathrm{rms}\,G\approx0.52$~\cite{Beenakker}.
In Fig.~\ref{FIG:universality}(b) we concentrate on n-doped graphene for step disorder
(upper panels of Fig.~\ref{FIG:disordermodels}) and show again
the universality of the SCF with respect to a wide range of parameters
characterizing edge roughness, ribbon length and width.
There is little dependence on the ribbon width $W$, confirming that the
observed effect is entirely due to the edges.

Currently there is not much experimental control over the edges of
nanoribbons. Considering GNRs with dense disorder, $d=\mathcal{O}(a)$,
the observed maximum of SCF decreases with increasing disorder density,
i.e.~decreasing $d$, as shown in the inset of Fig.~\ref{FIG:universality}(b).
We observe that for $d> 5a$ the SCF
are independent of the maximum height of the steps.
Moreover, we find that the maximum value of the SCF is retained for $d\gtrsim 5a$.
As an example, the system depicted in the upper right corner of Fig.~\ref{FIG:disordermodels}
shows spin conductance $\approx 0.4 e/4\pi$.
The finite spin conductance of GNRs predicted above,
and thus the existence of the edge state magnetism,
can be detected by measuring charge conductance, e.g. by attaching ferromagnetic leads
in a two- or four-probe measurement similar to
Ref.~\cite{vanWees}, with one lead being a zigzag GNR.

% {\it Conclusions.}
In conclusion, we have discussed the spin transport properties of
graphene nanoribbons.
We have shown that
an ideal GNR has zero spin conductance but nonzero spin Hall conductance.
Moreover, only GNRs with imperfect edges exhibit a nonzero spin conductance
The fluctuations of the spin conductance are universal with a maximum rms conductance
$\approx 0.4 e/4\pi$.
Thus, graphene nanoribbons can be used as an efficient
alternative to ferromagnetic leads, paving the way to
all-graphene spintronics devices.

We thank B.J.~van Wees, A.~Morpurgo and M.~Shiraishi
for discussions. I.A.,~M.W.,~S.B.~and K.R.~acknowledge financial support by DFG
(SFB689, GRK638) and D.T. by NSF NIRT grant
ECS-0506309, NSF NSEC grant EEC-425826 and the A.~v.~Humboldt
Foundation.


\begin{thebibliography}{10}
\bibitem{SCunit}
The unit of spin conductance is $e/4\pi$, in contrast to the spin-resolved charge conductance
$G_{\uparrow,\downarrow}$, which is $e^2/h$.
\end{thebibliography}
\end{document}